# Simulating the blood transfusion system in Kenya: Modelling methods and exploratory analyses

## Authors
Yiqi Tian[1], Bo Zeng[1]*, Jana MacLeod[2,3]*, Gatwiri Murithi[4]**, Cindy M Makanga[3]**, Hillary Barmasai[5], Linda Barnes[1,6], Rahul S. Bidanda[1], Tonny Ejilkon Epuu[3], Robert Kamu Kaburu[3], Tecla Chelagat[3], Jason Madan[7], Jennifer Makin[8], Alejandro Munoz-Valencia[9], Carolyne Njoki[10,11], Kevin Ochieng[4], Bernard Olayo[4], Jose Paiz[12], Kristina E. Rudd[13], Mark Yazer[14], Juan Carlos Puyana[15,16], Bopaya Bidanda[1], Jayant Rajgopal[1], Pratap Kumar[3]

*These authors contributed equally to this work
**These authors contributed equally to this work

## Affiliations
[1] Department of Industrial Engineering, University of Pittsburgh, Pittsburgh, PA, USA
[2] Kenyatta University College of Health Sciences School of Medicine, Nairobi, Kenya
[3] Institute of Healthcare Management, Strathmore University Business School, Nairobi, Kenya
[4] Center for Public Health and Development, Kisumu, Kenya
[5] Kenya Blood Transfusion and Transplant Service, Nairobi, Kenya
[6] Linda S. Barnes Consulting, Seattle, WA, USA
[7] Warwick Medical School, University of Warwick, UK
[8] Department of Obstetrics, Gynecology, and Reproductive Sciences, University of Pittsburgh Medical Center Magee Women's Hospital, Pittsburgh, PA, USA
[9] Department of Anesthesiology and Perioperative Medicine, University of Pittsburgh, Pittsburgh, PA, USA
[10] Department of Anesthesia, Aga Khan University Nairobi, Nairobi, Kenya
[11] Surgery Department, Egerton University, Egerton, Kenya
[12] Department of Surgery, University of Pittsburgh, Pittsburgh, PA, USA
[13] Center for Research, Investigation, and Systems Modelling of Acute Illness (CRISMA), Department of Critical Care Medicine, University of Pittsburgh, Pittsburgh, PA, USA
[14] Department of Pathology, University of Pittsburgh, Pittsburgh, PA, USA
[15] Departments of Surgery and Critical Care Medicine, University of Pittsburgh, Pittsburgh, PA, USA
[16] Department of Global Surgery, Royal College of Surgeons in Ireland, Dublin, Ireland

## Corresponding Author:
Dr. Pratap Kumar, MD, PhD
Institute of Healthcare Management
Strathmore University Business School
P.O. Box 59857-00200
Nairobi, Kenya
pkumar@strathmore.edu

**Short Title:** Kenya blood system simulation model

**Word count including abstract, tables, and figure titles:** 6,553

# Abstract

The process of collecting blood from donors and making it available for transfusion requires a complex series of operations involving multiple actors and different resources at each step. Ensuring hospitals receive adequate and safe blood for transfusion is a common challenge across low- and middle-income countries, but is rarely addressed from a system level. This paper presents the first use of discrete event simulation to study the blood system in Kenya and to explore the effect of variations and perturbations at different steps of the system on meeting blood demand at patient level. A process map of the Kenyan blood system was developed to capture critical steps from blood donation to transfusion using interviews with blood bank, hospital and laboratory personnel at four public hospitals across three counties in Kenya. The blood system was simulated starting with blood collection, a blood bank where blood is tested and stored before it is issued, a major hospital attached to the blood bank, and several smaller hospitals served by the same blood bank. Values for supply-side parameters were based mainly on expert opinion; demand-side parameters were based on data from blood requisitions made in hospital wards, and dispatch of blood from the hospital laboratory. Illustrative examples demonstrate how the model can be used to explore impacts of changes in blood collection (e.g., prioritising different donor types), blood demand (e.g., differing clinical case mix), and blood distribution (e.g., restocking strategies) on meeting demand at patient level. The model can reveal potential process impediments in the blood system and aid in choosing between alternate strategies or policies for improving blood collection, distribution or use. Such a systems approach allows for interventions at different steps in the blood continuum to be tested on blood availability for different patients presenting at diverse hospitals across the country.

# Introduction

Blood and blood products are critical to improving patient outcomes across a wide range of medical situations from emergencies to chronic illnesses. Maintaining a stable and cost-effective supply of blood products entails a complex series of events within the vein-to-vein system, beginning with blood donation and ending with blood transfusion. Variations of these events are dictated by the availability of blood donors, supply chain and facilities limitations, and regional or national regulations and customs. Given the multitude of places where interventions could be introduced to make improvements in the process, a holistic, high-level perspective of the entire system can be very helpful when attempting to streamline the efficiency of the blood donation and transfusion processes [1] [2].

Such a perspective is especially important in resource-constrained health systems where the mismatch between supply and demand is most pronounced [3]. The World Health Organization (WHO) reports that African countries annually collect only 5% of the global donated blood supply, despite constituting 14% of the world's population [4]. The WHO estimates that collection rates are as high as 53.0 units per 1,000 population in high-income countries; Kenya however is estimated to collect less than 10 units per 1,000 population annually, although recent increases in donation rates are reported [4],[5]. This paucity is further accentuated by a high prevalence of conditions where blood product administration is an integral component of clinical management. Sickle cell disease, malaria, gastrointestinal parasite infestation such as hookworm, and nutritional deficiencies are prevalent and often associated with clinically significant anaemia requiring blood transfusion, especially in children [6],[7]. Two out of five maternal deaths in Kenya are related to obstetric haemorrhage [8]–[11]; obstetric emergencies and trauma patients are disproportionately affected by blood supply deficiencies, as well as time factors because of limited infrastructure and poor referral networks [12], [13].

In Kenya, blood products for transfusion are collected and regulated by the Kenya Blood Transfusion and Transplant Service (KBTTS). The KBTTS structure consists of a national coordinating unit and six Regional Blood Transfusion centres (RBTCs) that collect, test, store, and distribute blood products, with an additional 43 satellite blood centres that collect blood but send it for testing at RBTCs [14]. The transfusion system has undergone significant recent shocks, starting with funding shortfalls [15], followed shortly thereafter by the COVID-19 pandemic that resulted in long-term school closures and therefore significant drop in blood collection (blood donation drives in Kenya and much of Sub-Saharan Africa target high school students aged 16 years and older) [16]-[18]. The Pathways for Innovation in Blood Transfusion Services in Kenya (PITS Kenya) study, funded through a BLOODSAFE program grant from the National Heart, Lung, and Blood Institute (NHLBI), aims to understand the blood system in Kenya and implement innovations to improve blood availability at the point-of-care [19]. The study focuses on three distinct counties in Kenya chosen for the variety of the challenges they face relating to their diverse location, socio-economic and cultural characteristics, disease prevalence and healthcare infrastructure.

Discrete event simulation (DES) models have been used to analyse and optimise the design and/or operation of systems in a wide range of domains, and have been recognized as a flexible and powerful tool for predicting the response of healthcare systems to change [20]. Their healthcare system applications encompass many key areas, including optimising hospital operation management, modelling disease progression for informed treatment decisions, evaluating screening programs (e.g. breast cancer screenings), and assessing health behaviour interventions [21]. These varied applications have had a significant impact, leading to improvements such as efficient nurse scheduling [22], optimising workflows in emergency departments [23], and strategizing vaccine distribution processes [24]. A recent review [25] identified 231 studies focused on DES modelling in healthcare. The majority of these were conducted in high-resource settings; few studies applied DES to LMIC healthcare settings, and none tackled the challenges of blood supply in LMICs.

One aim of the PITS Kenya study is to evaluate the complex blood continuum (steps spanning blood collection, processing, storage, delivery and use) in these three Kenyan counties using an *in silico* DES model of the entire system. DES models are representations of complex systems or processes utilising computer programs [26]. They have been successfully used in many application domains [27], including healthcare delivery [25], to gain insights into the system without directly modifying it or expending significant resources. This paper describes the design and construction of a DES model for the blood continuum in Kenya using three representative counties. Using illustrative examples, the paper demonstrates the potential of a simulation model to assist in understanding the obstacles in the blood continuum, and to guide effective policy or strategy decisions.

## Materials and Methods

From May 2020 through March 2023, a mixed-methods approach was employed to acquire data on the blood continuum in four locations (large hospitals which routinely collected blood from donors) within three Kenyan counties: Siaya, Turkana, and Nakuru (Nakuru County included two locations). The methods for the broader PITS Kenya research program have been previously published [19].

**Process Mapping**

A process map is a visual representation of workflow within a system that comprises different processes. The first step in the development of the DES model was a process mapping exercise, with the goal of understanding and describing the different processes in the system, their interrelationships, and the overall workflow. This allowed for the identification of key elements of the system configuration and their interactions within the system, which formed the basis for the DES model, and also informed the collection of appropriate data for constructing the model. In this context, the sequence of steps in the blood continuum may be divided into three broad stages: (1) collection, (2) processing and storage, and (3) delivery and use.

Process maps were generated, one for each location (n=4), that captured these three broad stages of the blood continuum. The process maps revealed both commonalities and differences across the counties. Commonalities included the sources of blood, which in all locations were blood drives, replacement donors and voluntary (drop-in) donors. Differences were observed, for example, in where blood was tested for transfusion-transmissible infections (TTIs); one location conducted on-site testing, while the other three sent samples to RBTCs for testing. The industrial engineers, in collaboration with the Kenya team, guided the development of a comprehensive draft process map, integrating processes from the different counties; colour-coding was used to distinguish unique processes specific to each county, thereby creating a visual representation of both shared and distinct practices.

The engineers presented the draft process map to laboratory and blood bank managers in all four locations. Small groups of blood continuum representative experts reviewed the map and provided specific feedback that ensured the inclusion of diverse perspectives and local insights [28]. The revised map was then presented in a blood system stakeholders' workshop to national, county, and community-level stakeholders, who provided the final validation of the processes utilised. This final version of the process map (Supplemental Figure 1) was adopted for constructing a DES model.

**Kenya Blood System Model**

As suggested recently by Jacobs et al. [2] this study takes a holistic approach to build a comprehensive model that captures the primary features of the entire blood continuum in Kenya. The primary elements of the model include blood collection processes and facilities (satellite collection centres or RBTCs) in the collection stage; transportation, testing, componentization and storage of blood in the processing stage; and distribution to hospital laboratories, clinical practices and patient types (e.g., emergency or non-emergency) in the delivery and use stage.

There are too many elements to be able to model and simulate every detail and variation of processes in the blood continuum. Therefore, the first step was to identify elements that would be impossible to model objectively (e.g., individual human behaviours, subjective judgments, rare events or those for which it would be difficult to obtain data), and unrepresentative features that were unique to a single location. Processes common across the study locations were prioritised in order to build a generic model which could be readily adapted to the different locations. This yielded a framework for a model that would be a reasonable representation of the system with sufficient details on the critical components, and at the same time, be feasible to simulate and analyse.

This model framework consists of a blood collection centre (satellite or RBTC) that is located within or alongside a large hospital ("Alpha hospital"), and a set of smaller, geographically distant hospitals ("Beta hospitals"), that depend on the same collection centre for their blood supply. This framework, illustrated in Figure 1, using one RBTC and three satellite collection centres and the Alpha and Beta hospitals they serve, represents a modular subsystem that is repeated to form the national blood system; the simulation in this study was conducted on a smaller unit highlighted by a yellow circle, comprising of one satellite blood collection centre, one Alpha hospital and three Beta hospitals. The different stages of the blood continuum are modelled as follows:

Figure 1. Blood collection and delivery subsystems

*Blood collection:* Blood is collected at collection centres using three different modes: blood drives, family replacement donors (FRD), and walk-in/volunteer donors. Blood drives are organised by the collection centre to take place at two broad types of locations: (a) institutions such as schools or military units, or (b) locations within the community such as town-centres or religious sites. FRD are relatives or friends of a patient requiring a blood transfusion, who are mobilised to donate by the patient or their caregivers. Volunteer donors are individuals presenting to the collection centre to give an undirected donation (not in response to any specific patient's needs). All donors are administered a health questionnaire followed by a preliminary physical screening to ensure their eligibility to donate blood as per the KBTTS guidelines. For various reasons such as risk factors or pre-existing conditions, not all donors who present to donate blood will pass this preliminary screening.

*Processing and storage:* Collected blood is tested for TTIs before it can be qualified for use and stored. The model does not currently involve componentization of blood units as this paper focuses on the use of whole blood. Samples from the satellite centres are sent to the nearest RBTC for testing, but only on specific days of the week to ensure there is a sufficient number of samples in each batch transported. Any TTI-positive blood is discarded, and only the qualified blood is stored at the collection centre. Collection centres have a finite amount of storage capacity for blood, both pre-qualified and qualified.

*Delivery and use:* The demand side of the model explores how blood is consumed at hospitals. Daily demand for blood is generated at each hospital by patients who fall into one of two categories based on timing of need: emergency and non-emergency. When this clinical need occurs, a request is sent to the blood collection centre where qualified blood is stored (this reflects current processes across all locations studied). The amount of time that elapses before the blood reaches the patient depends on the hospital setting; this is relatively short if it is an Alpha hospital (which is typically close to the collection centre), and longer if it is a Beta hospital (which is usually further away from a collection centre). Figure 2 depicts the operations of this subsystem schematically.

Figure 2. Blood subsystem operations

**Discrete Event Simulation (DES)**

DES modelling is one of the primary industrial engineering approaches for optimally designing a system/process, or for improving an existing one [27]. DES models have two specific features: (1) they are dynamic, modelling how a system evolves over time, and (2) they are stochastic, and explicitly capture and account for uncertainties in the system as it evolves. Simio® software was used to perform the DES modelling in this study [29]. The DES model was used to mimic the behaviour of the blood continuum over a 12-month study period. The system performance measure selected for this study was the percentage of patients who have their blood demand satisfied over the one-year period. Other metrics such as time to satisfy blood demand will be explored in future iterations.

**Simulation Inputs and Sources**

The values of the input parameters used in the simulation were obtained from various sources ranging from recorded data and observations, to estimates and opinions from those with domain expertise as described above. Tables 1 and 2 detail the input data used in the simulation for the blood collection & processing, and distribution & use stages of the blood continuum, respectively.

The simulation assumes that random arrivals of blood donors and patients are according to a Poisson process. Estimates of times that are random (transportation times, patient wait times) were based on a triangular distribution with three consensus estimates provided by experts: low, most likely and high. Binary random variables (test pass rates) were modelled with a Binomial distribution. Lastly, some random variables that could take on multiple discrete values (e.g., blood units per patient, FRD response rates) were modelled using discrete probabilities.

Data for average number of donors at blood drives and average number of family replacement donors mobilised per patient are based on review of blood drive records and interviews with blood bank and clinical staff in the study locations (Table 1). Patient blood need at an Alpha hospital is based on quantitative data from blood requisition records but similar data was not collected at Beta hospitals for this study (this is part of ongoing work). Thus in the simulation, the patient mix at a Beta hospital was assumed to be similar to an Alpha hospital but reduced to 30% of the value. Estimates of the range of maximum patient wait times for blood (for both emergency and non-emergency patients) can vary depending on the patient's specific condition, and therefore, the estimates used in the model (Table 2) were obtained in consultation with local study site clinicians.

Table 1: Blood collection and processing parameters

|  | Parameter | Probability Distribution | Source |
|---|---|---|---|
| COLLECTION | | | |
| Institutional Drives | every odd month; mean=100 /session | Poisson; $\lambda=100$ | Blood drive records, blood bank staff |
| Community Drives | every even month; mean=30 /session | Poisson; $\lambda=30$ | |
| Walk-In donors | Mon., Wed., Fri.: mean=1/day<br>Tue., Thu. mean=0.5/day | Poisson; $\lambda=1$<br>Poisson; $\lambda=0.5$ | |
| FRD response rates | | | |
| Donors per Alpha hospital patient: | 0=65%; 1=25%; 2=10% | Discrete; $p(0,1,2)=(0.65,0.25,0.1)$ | Blood bank staff, Clinical staff |
| Donors per Beta hospital patient: | 0=90%; 1=10% | Discrete; $p(0,1)=(0.9,0.1)$ | |
| ELIGIBILITY TESTING | | | |
| Donor pass rate (FRD) | 80% | Bernoulli; $p=0.8$ | Blood bank staff |
| Donor pass rate (other) | 90% | Bernoulli; $p=0.9$ | |
| TTI SCREENING | | | |
| Sample transport time (hours) | Every Tue., Thu. Satellite=5; RBTC=1 | - | Blood bank staff |
| Turn-around time estimates (hours) | low=4; likely=12; high=24 | Triangular ($a$=4, $m$=12, $b$=24) | |
| TTI test pass rate | FRD=90%; others=98% | Bernoulli; $p=0.9$ or $p=0.98$ | |

Table 2: Blood distribution and use parameters

|  | Parameter | Probability Distribution | Source |
|---|---|---|---|
| FOR ALPHA HOSPITALS | | | |
| Emergency patients | average of 1.3 per day | Poisson; $\lambda=1.3$ | Blood requisition records, clinical staff |
| Non-emergency patients | average of 4.7 per day | Poisson; $\lambda=4.7$ | |
| Blood units per patient | 1=35%; 2=50%; 3=15% | Discrete; $p(1,2,3)=$ (0.35,0.50,0.15) | |
| Transportation time estimates (hours) | low=1; likely=3; high=12 | Triangular ($a$=1, $m$=3, $b$=12) | Blood bank staff |
| Maximum emergency patient wait time (hours) | low=4; likely=24; high=36 | Triangular ($a$=4, $m$=24, $b$=36) | Clinical staff |
| Maximum non-emergency patient wait time (hours) | low=48; likely=72; high=168 | Triangular ($a$=48, $m$=72, $b$=168) | |
| FOR BETA HOSPITALS | | | |
| Emergency patients | 30% of Alpha hospital rate | Poisson; $\lambda=0.39$ | Clinical staff |
| Non-emergency patients | 30% of Alpha hospital rate | Poisson; $\lambda=1.441$ | |
| Blood units per patient | 1=35%; 2=50%; 3=15% | Discrete; $p(1,2,3)=(0.35,0.50,0.15)$ | |
| Transportation time estimates (hours) | low=12; likely=48; high=96 | Triangular ($a$=12, $m$=48, $b$=96) | Laboratory staff |
| Maximum emergency patient wait time (hours) | low=4; likely=24; high=36 | Triangular ($a$=4, $m$=24, $b$=36) | Clinical staff |
| Maximum non-emergency patient wait time (hours) | low=48; likely=72; high=168 | Triangular ($a$=48, $m$=72, $b$=168) | |

**Simulation Outputs**

A patient-centric measure was selected in this study to evaluate blood system performance: the percentage of patients who have their blood demand met. While the simulation can differentiate between fully and partially met (only some of the blood units requested are dispatched) demand, the results described here combine both into one measure of met demand. A three-month warm-up period was used in the simulation to allow the system to reach a steady state and to overcome any bias from the initial starting state of the system; all output data for this initial interval of time was discarded. The outputs for the following 12-month period were used to evaluate the system's performance. The simulation is repeated 100 times, and the outputs generated from each run are then averaged for analysis. Given the inherent variability in the system, the performance measure (percentage of met demand over a one-year period) from one run to another will vary. However, the probability distributions of inputs and the values of their associated parameters are constant from run to run. Thus, one would expect the performance measure from each run to vary randomly around the same mean value. In particular, with a sample size of 100, the confidence intervals (at the 95% level) for the mean performance measure were extremely tight and did not capture the variability in the performance across individual runs. Therefore, the median value for performance across the 100 replicates is reported along with the lower and upper quartiles.

## Results

Three experiments were chosen to illustrate the utility of the model. These examples were specifically chosen to illustrate the overarching "vein-to-vein" nature of the model, with experiments addressing the three different stages of the blood continuum: blood collection, processing & storage, and distribution & use.

EXPERIMENT 1: The first experiment simulates the impact of increasing blood supply in the system on meeting blood demand. Increased supply was simulated in two ways by increasing by roughly 50%, relative to baseline, either a) the total number of annual blood drives, or b) the number of FRDs mobilised per patient. The two changes represent two different ways in which blood supply is mobilised: the first is relatively unlinked to demand (blood drives are scheduled in advance and mostly conducted independent of hospital need for blood), while the second is directly linked to patients needing blood. In the simulation the changes were implemented by a) increasing the number of community drives by holding them once every month rather than once every other month; annual blood drives increased from 12 (6 institutional + 6 community; Table 1) to 18 (6 institutional + 12 community), or b) increasing FRD rates from p(0,1,2)=(65%,25%,10%) to p(0,1,2)=(45%,40%,15%) at the Alpha hospital, and from p(0,1)=(90%,1%) to p(0,1)=(85%,15%) at Beta hospitals.

Figure 3 displays, across the 100 runs, the percentage of patients whose blood order requests are fulfilled at baseline, and for each of the two simulation cases noted above; these are presented separately for the Alpha and Beta hospitals. As the figure shows, across all scenarios, Beta hospitals have a median met demand that is roughly 7 to 8 percentage points lower than the Alpha hospital: 21.5% (20.5-22.5%) vs 28.3% (27.4-29.2%). This is presumably because they are further away from the blood bank. For both Alpha and Beta hospitals there is a clear improvement in performance over the baseline with either scenario for increasing blood supply. This improvement in meeting demand is however much more pronounced with the increased blood supply from FRDs as compared to the increase from community blood drives: 35.6% vs 30.5% in Alpha and 27.6% vs 23.7% for beta hospitals, respectively.

Figure 3. Increased supply: median and quartiles of met demand for blood by hospital type

Abbreviations: BD, Blood Drive; FRD, Family Replacement Donor

Figure 4 also displays the results of Experiment 1, displayed by patient type (emergency vs. non-emergency). At baseline, non-emergency patients have a higher median rate of met demand compared to emergency patients (26.1%; IQR 25.4-26.4%) vs 21.9%; IQR 21.3-22.9%, respectively), likely because they can wait longer to receive blood. When the number of community blood drives is doubled, fulfilment of blood requests rises by approximately 2% for each patient type, to 28.1% (27.5-28.7%) for non-emergency and 24.6% (23.1-25.6%) for emergency patients. However, with doubling FRD rates, the outcome improves by nearly 7% for both patient types, to 32.7% (32.1-33.4%) for non-emergency patients and 28.6% (27.3-29.5%) for emergency patients.

Figure 4. Increased supply: median and quartiles of met demand for blood by patient type

Abbreviations: BD, Blood Drive; FRD, Family Replacement Donor.

EXPERIMENT 2: The second experiment focuses on the demand side for blood, and explores the change in met demand when the patient case mix (i.e., proportion of emergency vs. non-emergency patients, or E/NE ratio) changes from the baseline values. Although the patient mix is not something that can be directly controlled, the results of this experiment could provide guidance on anticipated blood need at different hospital locations depending upon local case mix. For the same average load of 6 patients per day at an Alpha hospital, the daily average E/NE ratio is changed from a baseline of 1.3/4.7 (Table 1) to 2.3/3.7. A contrasting scenario (E/NE ratio of 0.3/5.7) where there is a significantly larger share of non-emergency patients than in the baseline, was also studied. It was assumed the demand case mix for each Beta hospital is the same on average as at the Alpha hospital. A third scenario was also simulated where the patient mix is unchanged from the baseline, but some event (e.g., a road traffic accident with multiple severely injured patients) causes a sudden spike in the number of emergency patients at every hospital once per month.

Figure 5 displays the percentage of patients whose blood demand is satisfied, for the baseline and for each of the above three scenarios, and disaggregated between Alpha and Beta hospitals. The overall baseline pattern of improved performance at Alpha hospitals continues to be shown in all three scenarios. In each of the three scenarios there is no difference in median met demand relative to baseline. Therefore, assuming other model inputs and assumptions are constant, change in the proportion of patients with emergency and non-emergency blood needs does not appear to have a significant impact on the system.

Figure 5. Change in demand mix: median and quartiles of met demand for blood by hospital type

EXPERIMENT 3: The third experiment examines the effects of operational and policy changes in how blood is distributed to, and used in hospitals. The baseline model captures current operations of the Kenyan blood system where there is typically no blood stored at a hospital for planned use, and all blood is delivered to the hospital on demand from the blood bank. Given the time required to transport blood to the hospital from the blood bank, this could delay availability of blood, especially for patients who need blood emergently, and especially at a Beta hospital that is relatively distant from the blood bank serving it. Therefore, an alternative scenario was simulated in which periodic automatic replacement (PAR) is implemented at each hospital. In this scenario, a small inventory of blood (PAR amount) is maintained at the hospital. In this simulation, five units of tested blood are supplied from the blood bank at the beginning of every week. This blood is placed into storage at the hospital and made available to patients on a first-come, first-served basis. Two different restocking scenarios were explored, one where PAR amounts are maintained at both Alpha and Beta hospitals, and another where PAR amounts were maintained only at Beta hospitals (which are distant from blood banks).

Figure 6 shows that the system performance is virtually unchanged with both strategies, with no appreciable difference between them from baseline. However, the results in Figure 7, where the performance is separated out by type of hospital, are different and provide more insight. With the first strategy of restocking at both Alpha and Beta hospitals, met demand at the Alpha hospital deteriorates by a little under 3%, from 28.5% (27.4-29.2%) at baseline to 25.6% (25.1-26.6%), while met demand at Beta hospitals improves by about 4%, from 21.5% (20.5-22.5%) at baseline to 25.4% (24.6-26.1%). With the second strategy of restocking only at Beta hospitals, the drop in performance at the Alpha hospital is minimal (27.9%; 27.1-28.8%), but with modest improvement in performance at the Beta hospitals (22.9%; 22.0-23.9%). This outcome illustrates that in the "Refill Beta Hospital" scenario, the Alpha hospital performs worse because a portion of the blood supply is diverted to Beta hospitals. In the "Refill Both Hospitals" scenario, the Alpha hospital stores more blood and requests less from the central supply, which leads Beta hospitals to queue more frequently. This in turn deprioritizes the Alpha hospital's requests, further worsening its performance compared to the first strategy.

Figure 6. Changes in restocking policies: median and quartiles of overall met demand for blood

Figure 7. Changes in restocking policies: median and quartiles of met demand for blood by hospital type

An additional scenario was developed with a variation in the blood administration policy, where any blood in hospital storage (e.g., blood units that were delivered to the hospital but not transfused, possibly because the patient for whom blood was requested died before transfusion) is reserved only for emergency patients. Figure 8 shows results for the two different baselines (scenarios A, B), one with the current first-come first-served policy (met demand for non-emergency patients 26.1% (25.4-26.6%) vs. 21.9% (21.3-22.9%) for emergency patients) and one with priority for emergency patients (met demand for non-emergency patients 22.9% (22.2-23.9%) vs. 34.6% (33.5-35.9%) for emergency patients). A policy reserving any stored blood for emergency patients improves met demand for emergency patients by about 13%, but met demand for non-emergency patients decreases by about 3%. However, there is no difference in overall met demand for blood between the two different policies (around 25% in both cases), presumably because the number of emergency patients is only about a quarter of the number of non-emergency patients.

Figure 8. Restocking policies: median and quartiles of met demand by patient type

Abbreviations: E-Only, blood prioritised for emergency patients only

There is no impact on performance with a PAR strategy for either type of patient with the current first-come, first-served policy for access to stored blood (scenarios A, C, and E in Figure 8). However, implementing the policy of reserving stored blood only for emergency patients alongside a PAR strategy leads to additional increases in met demand for emergency patients (scenarios B, D and F in Figure 8), with varying reductions in met demand for non-emergency patients. In particular, the largest increase in meeting blood demand for emergency patients is with a PAR strategy at both Alpha and Beta hospitals as opposed to PAR at just the Beta hospitals. In patient numbers, with a baseline strategy where stored blood is reserved for emergency patients (scenario B), on average there were 312 emergency patients and 751 non-emergency patients with met demand (total of 1063 patients) in the simulation over one year. When this baseline is augmented with a PAR strategy at all hospitals (Scenario F), there are on average, 429 emergency and 627 non-emergency patients with met demand (total of 1056 patients). In other words, although the total number of patients with met demand is almost the same, there are on average 117 additional emergency patients who receive (potentially life-saving) blood, at the expense of 124 fewer non-emergency patients who might be compromised by not receiving the blood they need.

## Discussion

This novel study presents a comprehensive DES model of the blood transfusion system in Kenya, developed from data and insights across three diverse Kenyan counties, and demonstrates its usefulness in evaluating potential initiatives to improve the availability of safe blood for transfusion. This is the first attempt to simulate the entire blood supply chain from donation to transfusion at patient level in resource-constrained settings typical of most low- and middle-income country (LMIC) contexts. This comprehensive simulation model highlights the complexities and interdependencies of various stages within the blood transfusion continuum, and could provide policy makers with an analytic tool to assess various interventions in the blood system, helping to optimize blood availability for patients with conditions ranging from maternal emergencies to chronic diseases and cancer.

The few previous DES studies on blood systems in high-resource settings have typically focused on individual segments of the transfusion chain, such as the use of simulation to improve human resource allocation and management in blood collection systems [30], to optimise operational efficiency and service quality within a collection centre [31], or to address inventory strategies in blood centres, exploring ways to minimise wastage and shortage [32]. However, these prior studies do not encompass the complete blood transfusion process, including hospital processes, patient needs, and the potential interaction between patients and donors. While these studies provide valuable insights into parts of the overall process, they overlook the broader system and unique aspects of blood systems in LMICs like reliance on FRDs, limited testing capacity, diverse case mix driving blood demand, and other local socioeconomic, political, financial and clinical constraints affecting blood transfusion.

The DES model presented here offers a novel, unifying framework that captures key aspects across the entire blood transfusion continuum from donor accrual to patient transfusion. Each step within the system, from blood drives and donor eligibility to testing, storage, and distribution, involves numerous complexities and potential bottlenecks. By simulating these interconnected processes, the model can reveal critical points where interventions can yield the most significant improvements, offering a powerful tool to optimise the entire blood supply chain. One of the key features in this model is the differentiation between Alpha and Beta hospitals based on their proximity to blood banks. There are 49 blood collection centres in Kenya which collect and store blood from donors (of which six RBTCs also test blood for TTIs) [14]; however there are likely >1,000 hospitals that transfuse blood across Kenya [33]. Improving health outcomes across LMICs (e.g. reducing mortality due to obstetric haemorrhage) may require making blood available in distant transfusing facilities, with complex logistics related to collecting, testing and transporting blood that are included in the model.

Instead of using an outcome of increased blood supply like many previous studies, this study employs a patient-centric endpoint of the proportion of patient-level blood demand that is met, separated further along different patient types based on urgency of blood need. The selected metric directly reflects the effectiveness of the blood supply chain in meeting patient needs and provides a clear indicator of the impact of different interventions on patient outcomes. The model also has the capability to use various outcomes, such as blood unit level metrics, depending on the unique needs or preferences of patients or decision-makers. This flexibility allows researchers, healthcare administrators and policymakers to tailor the analysis to specific needs and priorities, ensuring that the model's insights are as relevant and actionable as possible.

There are two potential primary uses of this, or a similar, model for blood system stakeholders. First, the model could aid in understanding the sensitivity of the system to various input parameters that might not all be within direct control (e.g., the TTI pass rate, the mix of patients arriving at a particular hospital, or FRD response rates). This allows for a better understanding of potential impact or differences across contexts and would allow for more tailored allocation of human and material resources. Second, the model can be used to study the effects of planned or proposed interventions that could affect system behaviour (e.g., adding more blood drives, using faster modes of transportation for blood, sending blood samples for qualification testing more frequently, storing and testing blood at the point of use, etc.). Many of these interventions could be evaluated with changes only to the input data; the goal of the team in future work is to minimise the need to modify the simulation logic, and ensuring it is easily adaptable for various sites across Kenya.

The three experiments presented here highlight the potential usefulness of this model approach. First, Experiment 1 demonstrates why consideration of demand-linked supply (FRDs are explicitly linked to patients with blood need) is important. The experiment describes comparable changes to two different strategies in collecting blood, one linked to demand and the other not. Both strategies target donors in the same community: either relatives of patients, or other members of the same community targeted by drives in town centres (institutional drives typically target student populations). A 50% increase to current numbers in either strategy (recruitment of FRDs per patient or number of blood drives) results in differing numbers of donated units: adding community blood drives yields around 150 more qualified blood units annually, while increasing the FRD donation rate results in approximately 470 more units. While performance in the model is improved by maintaining a link between supply (blood donation) and demand (transfusion), future work will also need to compare costs and cost-effectiveness of either strategy. While the results do not advocate for any specific strategy in isolation, the experiment suggests that, based on the parameter set and assumptions in this model, strategies to maintain or increase demand-linked supply are important to consider, especially in smaller, remote hospitals where organising frequent blood drives might be challenging and less effective. In the real world in Kenya however, FRDs in smaller, remote hospitals, likely face challenges to reach distant blood collection centres to donate blood. Future work will explore combinations of strategies where collection strategies in Beta hospitals are paired with structural changes to the blood system (e.g. local collection and testing).

Experiment 2 demonstrates that the model system is not very sensitive to variations in emergency vs non-emergency case mix or sporadic increases in emergency blood need. This is possibly due to the low overall met demand rate (30%), meaning that the system is experiencing a blood shortage regardless of demand fluctuations. While low met demand rate at baseline is reflected in quantitative data collected in the study (manuscript in preparation), we also ran all the experiments described here after artificially increasing blood supply. There were no qualitative differences to the findings presented here in a paradigm where baseline met demand rate is higher (data not shown). When a patient is discharged before receiving their requested blood, their unused request is reallocated to the next patient, and this rollover mechanism helps maintain stable fulfillment rates at the hospital level, even when the patient mix changes. This finding emphasises that the rate of meeting blood demand may be less variable from setting-to-setting based on case mix, and focus on other elements of the system may have a greater impact in informing potential future planning and interventions.

An interesting finding of the Experiment 3 was that a regular restock of blood to hospital labs could help improve meeting blood demand for patients in Beta hospitals without significantly compromising the overall met demand rate. Combining this restock strategy with reserving some blood for emergency use could significantly improve the met demand rate for emergency patients, while compromising care for non-emergency patients. The simulation therefore reveals the policy decisions that may be necessary, and their impact, in strategies that demonstrate great potential for ensuring a steady availability of blood for both routine and emergency needs, and reducing delays in critical situations.

There are several limitations to this study. First, the results presented here are not meant to be definitive, nor are they meant to be prescriptive. Rather, the goal of this project is to provide illustrative examples of how a realistic, "vein-to-vein" simulation model of the blood system can be used to gain insights into a complex system, and how it can serve to evaluate various what-if scenarios and decision alternatives. Some aspects of the blood continuum have been simplified for modelling purposes (e.g., lumping all patients into two broad categories based on urgency of blood need, assuming reasonable probability distributions for random factors, and focusing on blood as a single product while ignoring blood typing and specific blood components such as plasma or platelets). Therefore, further assessment, both within and outside of the model, must be conducted prior to implementing the results of these simulations into practice. Second, while the model presented here uses actual data wherever such data are available (e.g., on the demand side), and direct feedback from practitioners in the field where reliable data is unavailable, it does not perfectly reflect the reality of the entire system in every possible context. Thus, while the results presented are internally valid and externally informative, the specific estimated findings may not perfectly represent real-world results. Third, the model presented here is stochastic, in that it captures fluctuations in flows and events over time, but is not probabilistic, in that it does not reflect uncertainty in parameters such as mean daily events. This distinction is sometimes referred to as 'first order' versus 'second order' uncertainty [34]. Further development of the model to include second-order uncertainty would allow for probabilistic sensitivity analysis [35] that could provide additional value to policymakers by quantifying uncertainty around recommendations and identifying where further data collection would be most valuable. However, the data required for probabilistic modelling of this blood continuum are not currently available.

In conclusion, this paper presents the methods and illustrative results of a DES simulation model of the blood continuum in Kenya. By considering the entire blood continuum from "vein to vein," the use of both quantitative and qualitative real-world data to inform model inputs, inclusion of a patient-centric outcome, and an incorporating diverse LMIC blood system contexts, this model provides a new and insightful approach to understanding and improving global health systems. Future research using this model, or incorporating approaches used in developing this model, could improve the ability of health system leaders to make operational, policy and funding decisions to increase the availability of safe blood for transfusion at the point-of-care across LMICs.

## Acknowledgments

The BLOODSAFE program is supported by research grants from the National Heart, Lung, and Blood Institute (NHLBI grants UG3HL151595, UG3HL151599, UG3HL152189, and U24HL151541). The primary investigators of the BLOODSAFE program scientific projects are Lucy Asamoah-Akuoko, Yvonne Dei-Adomakoh, Mina C. Hosseinipour, Pratap Kumar, Bridon m'baya, and Juan Carlos Puyana. The data coordinating center primary investigator is Cavan Reilly, the steering committee chairperson is Meghan Delaney. The authors would like to thank members of the NHLBI Observational Study Monitoring Board (OSMB)/Data and Safety Monitoring Board (DSMB) for their review of the protocols and study progress: E.L. Murphy (Chair), E. Bukusi, H.A. Hume, O. Ogedegbe, S. Person and A.E. Mast (ad hoc).

## Conflicts of Interest
PK is a director in Health-E-Net Limited, which has commercial interests in the PaperEMR® technology used to capture demand data on blood transfusion. LSB is a consultant to the blood and biotherapies sector including the Association for the Advancement of Blood and Biotherapies (AABB). None of the opinions or views expressed in this manuscript obligate, bind, or otherwise commit the AABB. Research reported in this publication was supported by the National Heart, Lung, And Blood Institute of the National Institutes of Health under Award Number UGSHL151595. The content is solely the responsibility of the authors and does not necessarily represent the official views of the National Institutes of Health. No other commercial interests are reported for this manuscript. All other authors have no competing interests.

# References

[1] M. Delaney *et al.*, "The BLOODSAFE program: Building the future of access to safe blood in Sub-Saharan Africa," *Transfusion*, vol. 62, no. 11, pp. 2282–2290, 2022, doi: 10.1111/trf.17091.

[2] J. W. Jacobs et al., "Ensuring a Safe and Sufficient Global Blood Supply," *New England Journal of Medicine*, vol. 0, no. 0, pp. 0–0, 2024, doi: 10.1056/NEJMp2403596.

[3] C. F. Nathan Roberts, Sarah James, Meghan Delaney, "The global need and availability of blood products: a modelling study," *Lancet Haematol.*, vol. 6, no. 12, pp. e606–e615, 2019.

[4] WHO, "Global status report on blood safety and availability 2021," Geneva, 2022. [Online]. Available: https://www.who.int/publications/i/item/9789240051683.

[5] A. J. D.-N. J.B. Tapko, Paul Mainuka, "Status of Blood Safety in the WHO African Region: Report of the 2006 Survey," 2009. [Online]. Available: https://iris.who.int/handle/10665/364673.

[6] H. M. Nabwera *et al.*, "Pediatric blood transfusion practices at a regional referral hospital in Kenya," *Transfusion*, vol. 56, no. 11, pp. 2732–2738, 2016, doi: 10.1111/trf.13774.

[7] A. Esoh, K.; Wonkam-Tingang, E.; Wonkam, "Sickle cell disease in Sub-Saharan Africa: transferable strategies for prevention and care," *Lancet Haematol.*, vol. 8, no. 10, pp. e744–e755, 2021.

[8] M. B. Holcomb JB, Wade CE; Trauma Outcomes Group; Brasel KJ, Vercruysse G, MacLeod J, Dutton RP, Hess JR, Duchesne JC, McSwain NE, Muskat P, Johannigamn J, Cryer HM, Tillou A, Cohen MJ, Pittet JF, Knudson P, De Moya MA, Schreiber MA, Tieu B, Brundage S, Napolit, "Defining present blood component transfusion practices in trauma patients: papers from the Trauma Outcomes Group," *J. Trauma-Injury Infect. Crit. Care*, vol. 71, no. 2 Suppl 3, pp. S315–S317, 2011, doi: https://doi.org/10.1097/TA.0b013e318227ed13.

[9] P. K. Christoph Massoth, Maximilian Wenk, Patrick Meybohm, "Coagulation management and transfusion in massive postpartum hemorrhage," *Curr. Opin. Anaesthesiol.*, vol. 36, no. 3, pp. 281–287, 2023, doi: 10.1097/aco.0000000000001258.

[10] P. K. Edelson *et al.*, "Maternal anemia is associated with adverse maternal and neonatal outcomes in Mbarara, Uganda," *J. Matern. Neonatal Med.*, vol. 36, no. 1, p., 2023, doi: 10.1080/14767058.2023.2190834.

[11] N. Muthigani, W., Ameh, C. A., Godia, P. M., Mgamb, E., Maua, J., Okoro, D., Smith, H., Mathai, M., van den Broek, "Saving Mothers' Lives: Confidential Inquiry into Maternal Deaths in Kenya, 2017 First Report," 2017.

[12] B. B. Masaba, J. K. Moturi, J. Taiswa, and R. M. Mmusi-Phetoe, "Devolution of healthcare system in Kenya: progress and challenges," *Public Health*, vol. 189, pp. 135–140, 2020, doi: 10.1016/j.puhe.2020.10.001.

[13] L. Yang *et al.*, "Evaluation of blood product transfusion therapies in acute injury care in low- and middle-income countries: a systematic review," *Injury*, vol. 51, no. 7, pp. 1468–1476, 2020, doi: 10.1016/j.injury.2020.05.007.

[14] Kenya Tissue and Transplant Authority, "Kenya Donation Centers," 2024. https://www.ktta.go.ke/donation-centers/ (accessed Jul. 02, 2024).

[15] P. Adepoju, "Blood transfusion in Kenya faces an uncertain future," *The Lancet*, vol. 394, no. 10203, pp. 997-998, Sep. 21, 2019, doi: 10.1016/S0140-6736(19)32140-3.

[16] J. O. Okuthe, E. W. Muitta, and A. O. Odongo, "Determinants of blood donation among selected tertiary college students in Homa Bay County Kenya," *International Journal of Community Medicine and Public Health*, vol. 9, no. 3, pp. 1250–1256, 2022, doi: 10.18203/2394-6040.ijcmph20220682.

[17] M. Akulume, A. N. Kisakye, F. R. Nankya, and S. N. Kiwanuka, "Predicting intention to donate

blood among secondary school students in Eastern Uganda: An application of the theory of planned behavior," *medRxiv*, 2024.06.20.24309241, doi: 10.1101/2024.06.20.24309241.

[18] S. E. Njolomole, B. M'baya, G. Mandere, E. Storey, A. Jenny, T. Chiwindo, F. Nyangu, D. Walker, and A. S. Muula, "Strategies to meet blood demand for transfusions during the COVID-19 pandemic: lessons learnt from a large central hospital in Malawi," Annals of Blood, vol. 8, no. 0, 2022.

[19] A. Munoz-Valencia *et al.*, "Protocol: identifying policy, system, and environment change interventions to enhance availability of blood for transfusion in Kenya, a mixed-methods study," *BMC Health Serv. Res.*, vol. 23, no. 1, pp. 1–12, 2023, doi: 10.1186/s12913-023-09936-0.

[20] J. Karnon, J. Stahl, A. Brennan, J. J. Caro, J. Mar, and J. Möller, "Modeling using discrete event simulation: A report of the ISPOR-SMDM modeling good research practices task force-4," *Value Heal.*, vol. 15, no. 6, pp. 821–827, 2012, doi: 10.1016/j.jval.2012.04.013.

[21] X. Zhang, "Application of discrete event simulation in health care: A systematic review," *BMC Health Serv. Res.*, vol. 18, no. 1, pp. 1–11, 2018, doi: 10.1186/s12913-018-3456-4.

[22] S. M. Qureshi, N. Purdy, A. Mohani, and W. P. Neumann, "Predicting the effect of nurse–patient ratio on nurse workload and care quality using discrete event simulation," *J. Nurs. Manag.*, vol. 27, no. 5, pp. 971–980, 2019, doi: 10.1111/jonm.12757.

[23] B. Easter, N. Houshiarian, D. Pati, and J. L. Wiler, "Designing efficient emergency departments: Discrete event simulation of internal-waiting areas and split flow sorting," *Am. J. Emerg. Med.*, vol. 37, no. 12, pp. 2186–2193, 2019, doi: 10.1016/j.ajem.2019.03.017.

[24] F. Sala, G. D'Urso, and C. Giardini, "Discrete-event simulation study of a COVID-19 mass vaccination centre," *Int. J. Med. Inform.*, vol. 170, no. November 2022, p. 104940, 2023, doi: 10.1016/j.ijmedinf.2022.104940.

[25] J. I. Vázquez-Serrano, R. E. Peimbert-García, and L. E. Cárdenas-Barrón, "Discrete-event simulation modeling in healthcare: A comprehensive review," *Int. J. Environ. Res. Public Health*, vol. 18, no. 22, 2021, doi: 10.3390/ijerph182212262.

[26] D. Kemp, *Discrete-event Simulation: Modeling, Programming, and Analysis*, vol. 52, no. 3. 2003.

[27] S. Robinson, "Discrete-event simulation: From the pioneers to the present, what next?," *J. Oper. Res. Soc.*, vol. 56, no. 6, pp. 619–629, 2005, doi: 10.1057/palgrave.jors.2601864.

[28] B. Bidanda *et al.*, "Researchers , Pathologists , and OB / GYNs to solve Wicked Health Systems Challenges : Takeaways from a Multi-Disciplinary Team Affiliations :," *Proc. Inst. Ind. Syst. Eng. Annu. Conf. Expo 2022*, vol. 2, pp. 1208–1214, 2022.

[29] C. D. Pegden, "Simio: A new simulation system based on intelligent objects," *Proc. - Winter Simul. Conf.*, pp. 2293–2300, 2007, doi: 10.1109/WSC.2007.4419867.

[30] E. Alfonso, X. Xie, V. Augusto, and O. Garraud, "Modelling and simulation of blood collection systems: Improvement of human resources allocation for better cost-effectiveness and reduction of candidate donor abandonment," *Vox Sang.*, vol. 104, no. 3, pp. 225–233, 2013, doi: 10.1111/vox.12001.

[31] M. Doneda, S. Yalçındağ, I. Marques, and E. Lanzarone, "A discrete-event simulation model for analysing and improving operations in a blood donation centre," *Vox Sang.*, vol. 116, no. 10, pp. 1060–1075, 2021, doi: 10.1111/vox.13111.

[32] F. Baesler, M. Nemeth, C. Martínez, and A. Bastías, "Analysis of inventory strategies for blood components in a regional blood center using process simulation," *Transfusion*, vol. 54, no. 2, pp. 323–330, 2014, doi: 10.1111/trf.12287.

[33] E. Mumo, N. O. Agutu, A. K. Moturi, A. Cherono, S. K. Muchiri, R. W. Snow, and V. A. Alegana, "Geographic accessibility and hospital competition for emergency blood transfusion services in Bungoma, Western Kenya," *International Journal of Health Geographics*, vol. 22, no. 6, 2023.

[34] A. H. Briggs, M. C. Weinstein, E. A. L. Fenwick, J. Karnon, M. J. Sculpher, and A. D. Paltiel, "Model parameter estimation and uncertainty: A report of the ISPOR-SMDM modeling good research practices task force-6," *Value Heal.*, vol. 15, no. 6, pp. 835–842, 2012, doi: 10.1016/j.jval.2012.04.014.

[35] G. Baio and A. P. Dawid, "Probabilistic Sensitivity Analysis in Health Economics Probabilistic Sensitivity Analysis in Health Economics," *Stat. Method Med. Res.*, vol. 6, no. 615–34, pp. 1–20, 2011.

# Figures

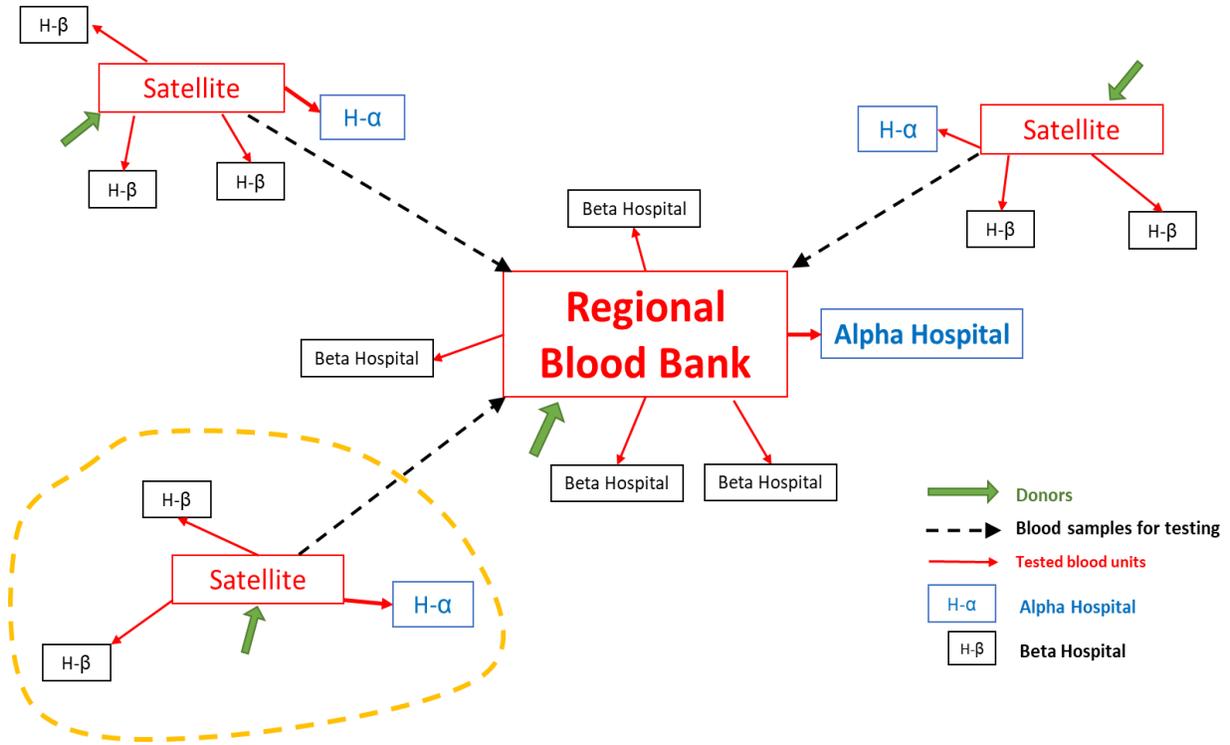

Figure 1. Blood collection and delivery subsystems

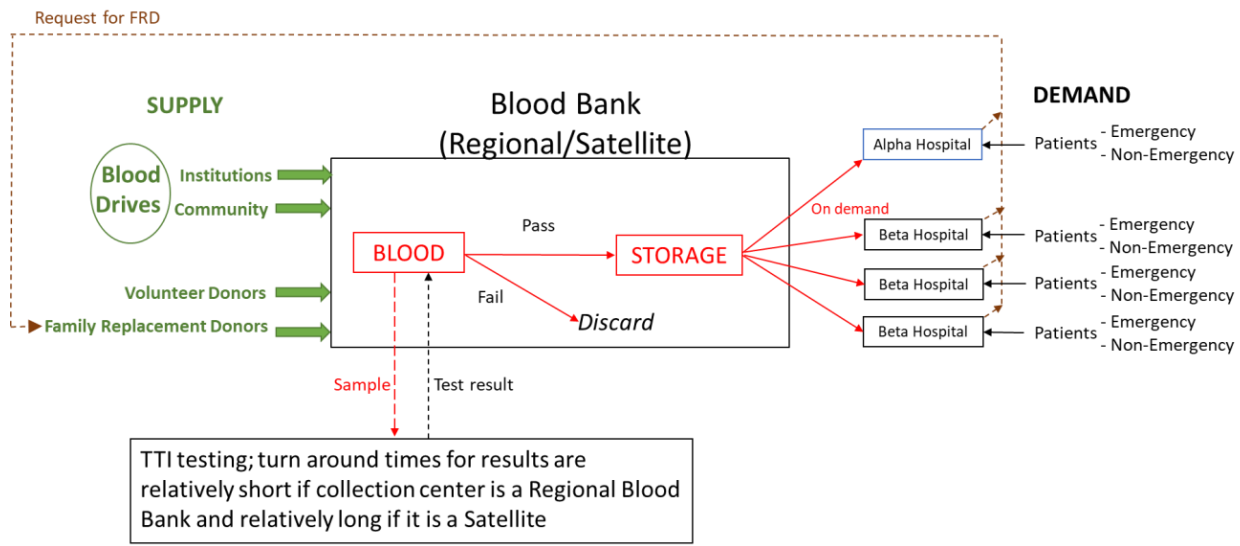

Figure 2. Blood subsystem operations

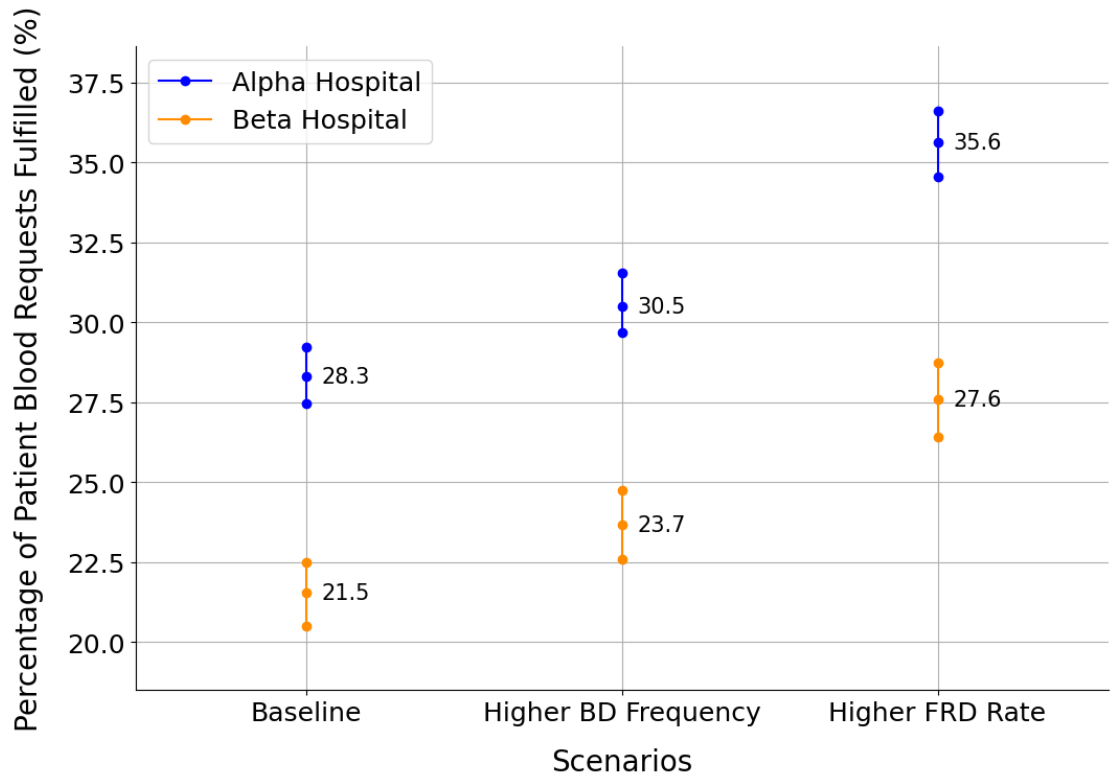

Figure 3. Increased supply: median and quartiles of met demand for blood by hospital type

Abbreviations: BD, Blood Drive; FRD, Family Replacement Donor.

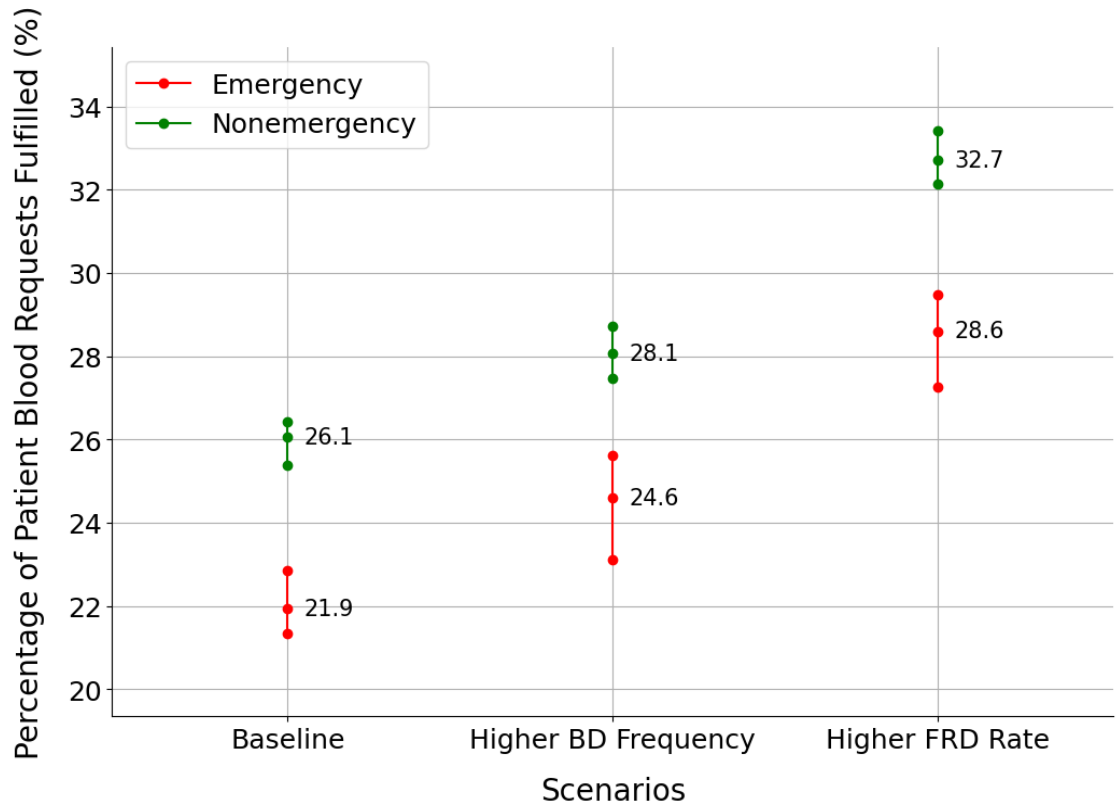

Figure 4. Increased supply: median and quartiles of met demand for blood by patient type

Abbreviations: BD, Blood Drive; FRD, Family Replacement Donor.

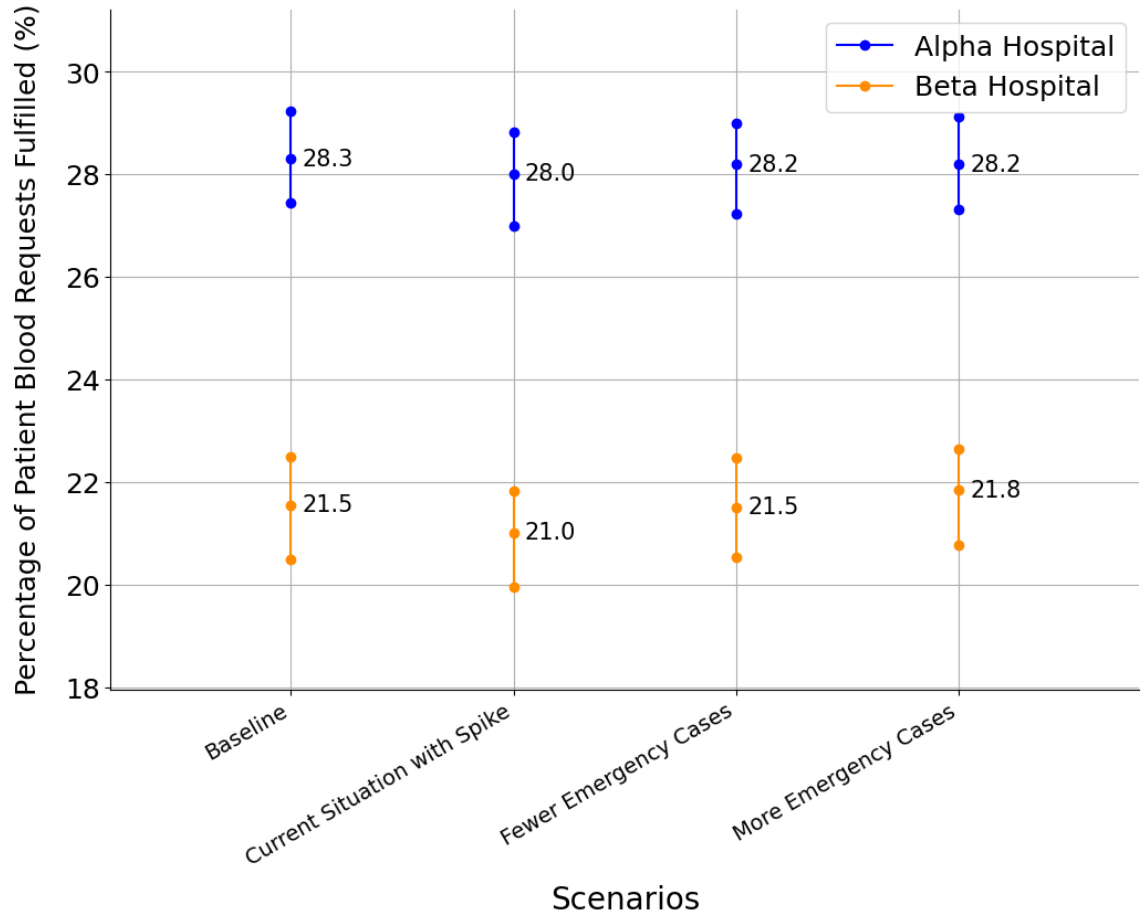

Figure 5. Change in demand mix: median and quartiles of met demand for blood by hospital type

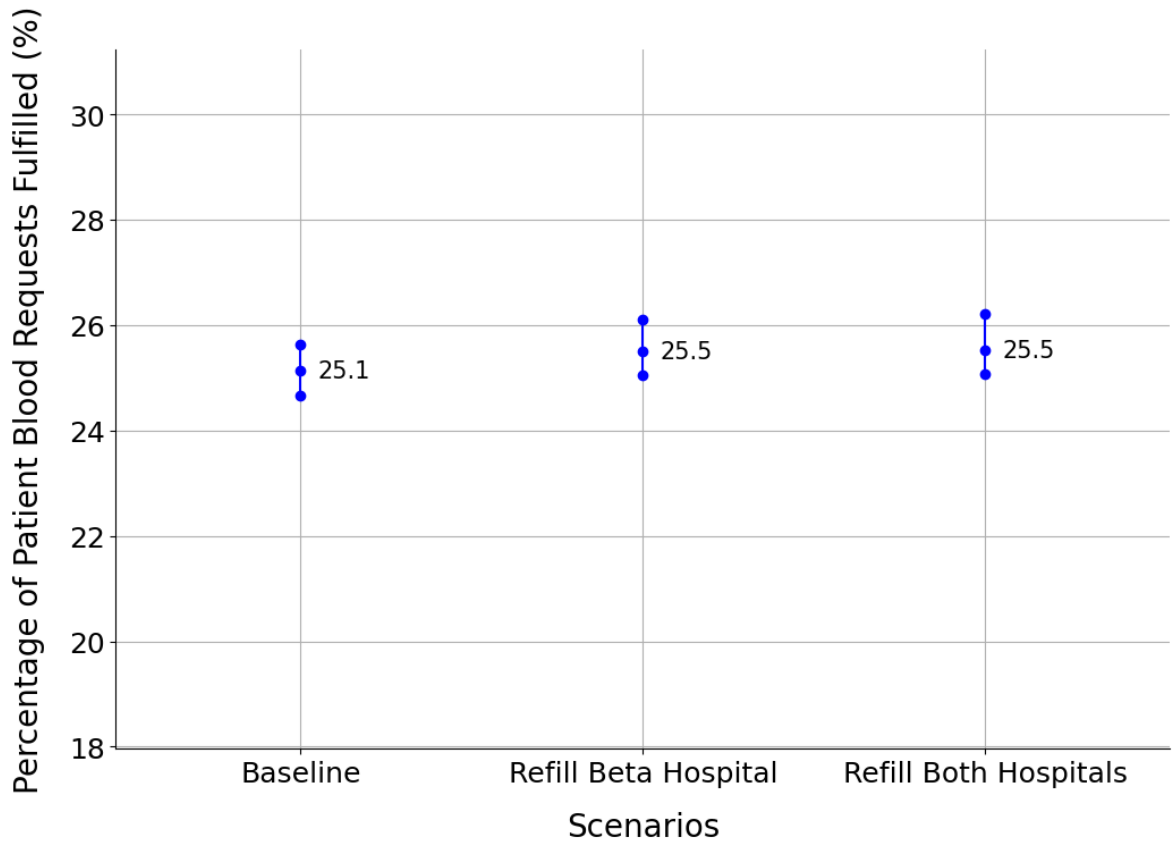

Figure 6. Changes in restocking policies: median and quartiles of overall met demand for blood

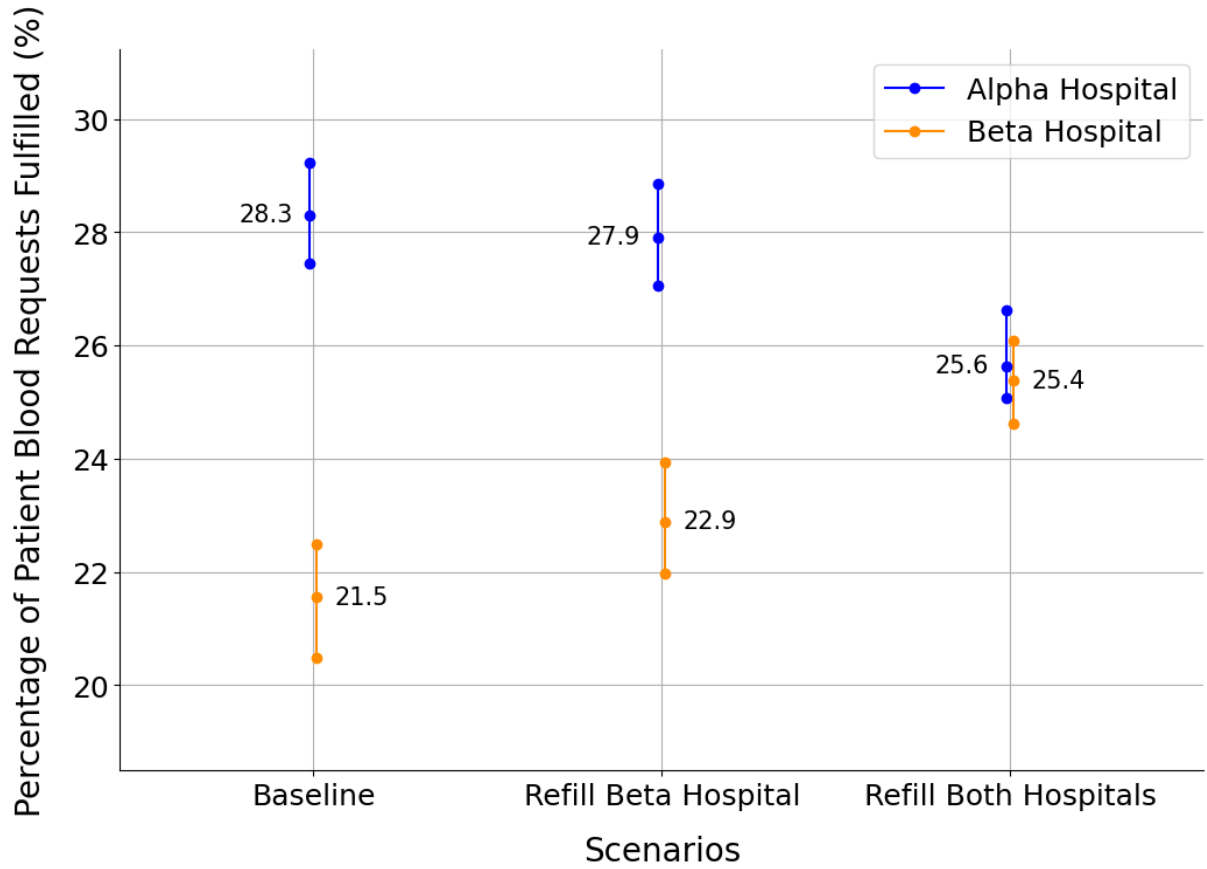

Figure 7. Changes in restocking policies: median and quartiles of met demand for blood by hospital type

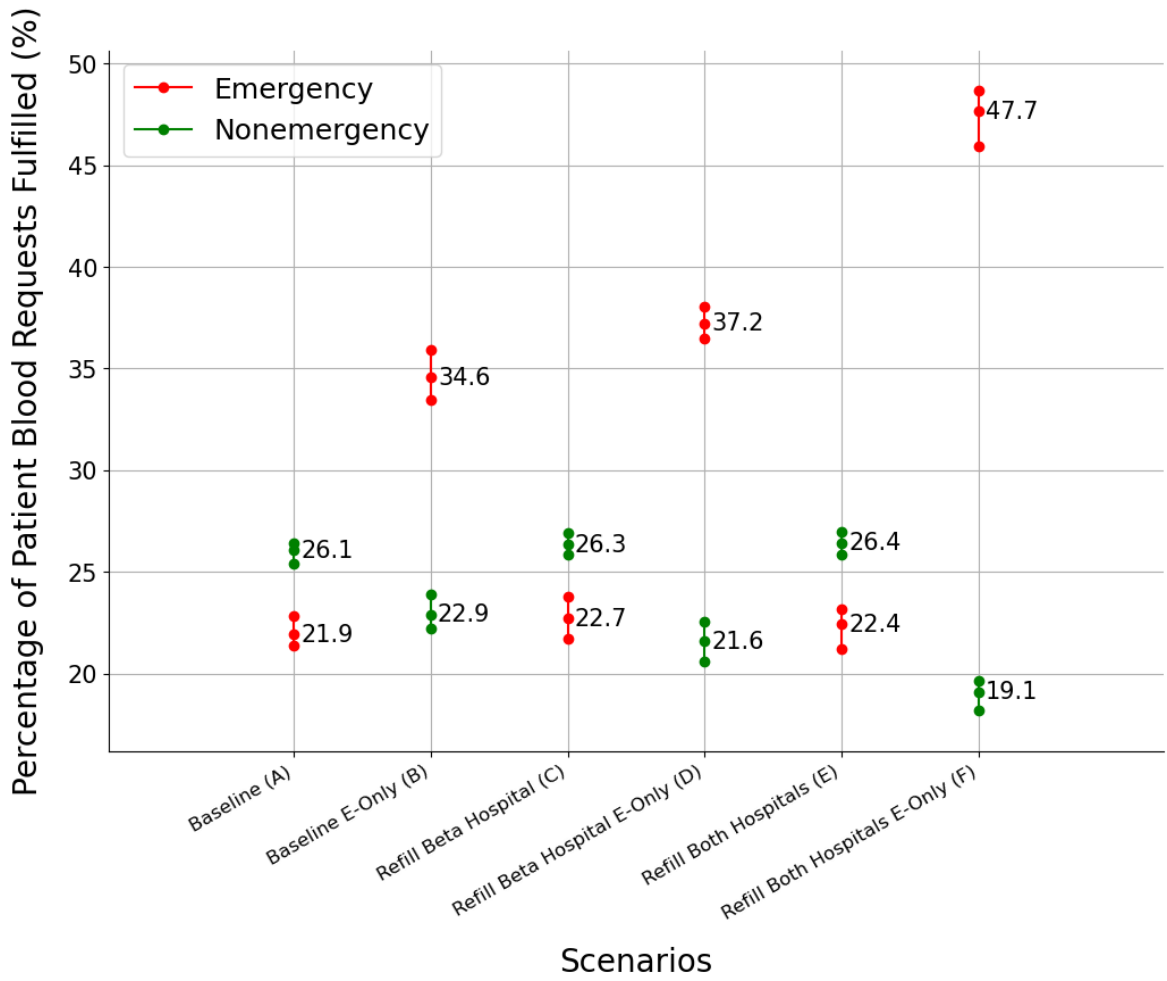

Figure 8. Restocking policies: median and quartiles of met demand by patient type

Abbreviations: E-Only, blood prioritised for emergency patients only

## Supplemental Figure

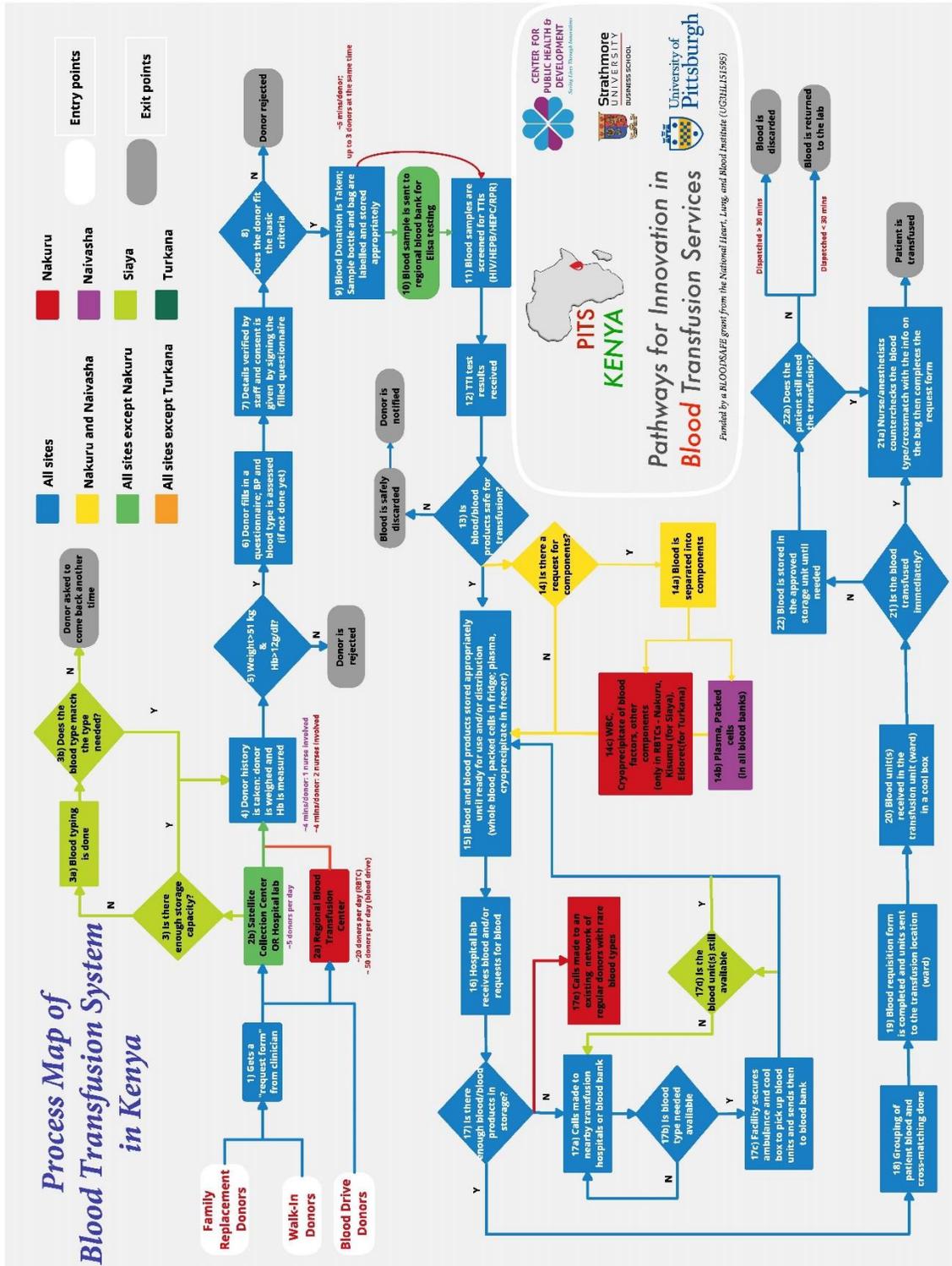

Supplemental Figure 1. Kenya Blood System Process Map